\documentstyle[prb,aps,preprint]{revtex}
\begin{document}
\draft
\title{The Model of the Theory of the Quantum Brain Dynamics 
can be cast on the Heisenberg Spin Hamiltonian}
\author{Tadafumi Ohsaku}
\address{Research Center for Nuclear Physics, Osaka University, Ibaraki, Osaka, Japan.}

\date{\today}
\maketitle

In this note, we show that the model of the quantum brain dynamics
can be cast on a kind of the Heisenberg spin Hamiltonian.
Therefore, we would like to emphasize that {\it the quantum dynamics of brain should be understood 
by the physics of quantum spin systems}. 

\vspace{10mm}

Until now, various models were proposed for the theory of brain.
All of these theories have to describe the mechanism of memory and retrieval.
The quantum field theory of brain was first established by Umezawa et al.~[1,2,3], 
and was developed by several groups~[4,5]. 
In the theory of Umezawa, he introduced a model Hamiltonian of brain. 
The model Hamiltonian ( so called Takahashi-Umezawa Hamiltonian ) has a symmetry, 
for instance, $SU(2)$ symmetry, and the symmetry is spontaneously broken. 
Umezawa speculated that the generation of the Nambu-Goldstone ( NG ) mode 
in brain corresponds to a memory-storage, 
while an excitation of the NG mode is a recalling of the memory. 
He constructed the model by two dynamical degrees of freedom: 
one is spin ( they called it {\it corticon} ),
and other is photon ( they called it {\it stuarton} ). 

Jibu et al. gave an interpretation of the spin variable in the Takahashi-Umezawa Hamltonian. 
They regarded the spin as the electric dipole field generated by water molecules in brain~[4]. 
Jibu et al. introduced a Hamiltonian of the electromagnetic field coupled with the electric dipole field,
and showed that the Hamiltonian is equivalent to the Takahashi-Umezawa Hamiltonian. 
Here we want to show that the Heisenberg spin Hamiltonian will be derived from the theory of Jibu et al.
We can write the essential part of the model of Jibu as follows:
\begin{eqnarray}
{\cal L} &=& -\frac{1}{4}F_{\mu\nu}F^{\mu\nu}-G{\bf S}\cdot{\bf A},
\end{eqnarray}
where, $F_{\mu\nu}=\partial_{\mu}A_{\nu}-\partial_{\nu}A_{\mu}$ and $G$ is the coupling constant
of the interaction of ${\bf S}$ and ${\bf A}$.
In Eq. (1), we wrote the electric dipole field as a spin variable ${\bf S}$.
Thus, after integrating out the electromagnetic field by using the photon propagator, 
we yield the Heisenberg Hamiltonian of the spin system:
\begin{eqnarray}
H_{spin} &=& -\sum_{i,j}J({\bf x}_{i},{\bf x}_{j}){\bf S}({\bf x}_{i})\cdot{\bf S}({\bf x}_{j}), \\
J({\bf x}_{i},{\bf x}_{j}) &=& \int\frac{d^{3}k}{(2\pi)^{3}}\frac{e^{i{\bf k}\cdot({\bf x}_{i}-{\bf x}_{j})}}{{\bf k}^{2}+m_{ph}}. 
\end{eqnarray}
Here, $i$ and $j$ running the number of the electric dipoles in brain, and $m_{ph}$ is the photon mass
generated by a possible charge shielding effect. $J$ is given by the photon propagator, 
and the photon propagator should include not only the charge shielding effect 
but also other possible environmental effects in brain.
Here we negrect the energy dependence of the photon propagator, 
and $J$ depends only on the spatial coordinates ${\bf x}_{i}$ and ${\bf x}_{j}$.
 
Therefore, based on our discussion given above, 
we regard brain is a kind of spin system and the NG boson of brain is interpreted as a magnon. 
Here we would like to make an argument that, 
{\it the dynamics in brain should be described by the theory of magnetism, 
especially by the theory of quantum spin system}. 
After we obtain this viewpoint, we can employ various powerful methods 
in condensed matter physics to study brain.
When the spin system is under a broken-symmetry phase, the system stores a memory, 
and the excitation of a magnon mode is the recalling of the memory.
For example, the thermodynamics of brain can also be studied by the method of the spin system.
The effect of the thermal noise can be incorpolated into the theory of brain. 
The Neel temperature or Curie temperature are interpreted as the temperatures of brain 
at which the disappearances of the memories might be happend. 
It is interesting for us to examine what character brain has; 
ferromagnetic, antiferromagnetic or other types of order.
It is well-known fact that the dispersion relation of the magnon is different between 
in the ferromagnetic and antiferromagnetic cases. 
If it is possible for us to measure a thermodynamic quantity of the brain,
we can determine such kind of character of brain. 
We can introduce the characteristic temperature of the brain
$T_{B}$ ( brain temperature ) in our theory.   
We should also pay attention to the effect of the dimensionality of brain.
The difference of the dimensionality of the system reflects the geometrical structure of the
spin model, the dependence on spatial coordinates in the coupling $J$,
and the physics of the NG mode. 
Various application of the methods of the spin system and the theory of statical/dynamical phase transition
can be employed to study the physics of brain.
Vitiello et al. argued that brain is an open system, and a consideration of the dissipation
is very important for understanding the mechanism of memorization and recalling~[5]. 
The effect can also easily be included in our spin Hamiltonian.


\begin{references}



\item 
L. H. Ricciardi and H. Umezawa, Kibernetik {\bf 4}, 44 (1967).
\item
C. I. J. M. Stuart, Y. Takahashi and H. Umezawa, J. Theor. Biol. {\bf 71}, 606 (1978), 
Found. Phys. {\bf 9}, 301 (1979).
\item
H. Umezawa, {\it Advanced Field Theory} (The American Institute of Physics Press, New York, 1993). 
\item
M. Jibu, K. H. Pribram and K. Yasue, Int. J. Mod. Phys. {\bf B10}, 1735 (1996).
\item 
E. Alfinito and G. Vittielo, quant-ph/0006065, quant-ph/0006066, quant-ph/0002014.



\end{references}
\end{document}